\newcommand{\option}{\iffalse}

\newcommand{\myTitle} {
On the possibility of measuring the polarization of a ${}^3\rm{He}$ beam at EIC by the HJET polarimeter
}
\newcommand{\myAbstract}{
The requirements for hadron polarimetry at the future Electron Ion Collider (EIC) include measurements of the absolute helion (${}^3\text{He}$, $h$) beam polarization with systematic uncertainties better than $\sigma^\text{syst}_P/P\lesssim1\%$. Here, we consider a possibility to utilize the Polarized Atomic Hydrogen Gas Jet Target (HJET) for precision measurement of polarization of the $\sim$100\,GeV/n helion beam. HJET, which serves to determine absolute proton beam polarization at the Relativistic Heavy Ion Collider, provides the accuracy of about $\delta^\text{syst}P/P\sim0.5\%$. Potential problems for adapting the HJET method  for the EIC helion beam include ({\em i}) necessity to know the ratio of $p^\uparrow{h}$ and $h^\uparrow{p}$ analyzing powers $A_\text{N}^{ph}(t)/A_\text{N}^{hp}(t)$ with high precision, ({\em ii}) possible beam ${}^3\text{He}$ breakup, and ({\em iii}) operation in a 10 ns bunch spacing beam. Preliminary results of an analysis discussed here indicate that the listed problems can be overcome and the helion beam absolute polarization can be measured by HJET with the required accuracy.
}
\newcommand{\myKeywords}{
  Hadron Polarimetry; He3 beam at EIC; Analyzing power; Breakup Corrections;
}                   

\option

\documentclass{jps-cp}>
\usepackage{txfonts} 
\title{\myTitle}

\inst{
$^{1}$Brookhaven National Laboratory, Upton, New York 11973, USA \\
$^{2}$School of Mathematics, Trinity College Dublin, Dublin 2, Ireland
}
\email{poblaguev@bnl.gov}
\recdate{January 15, 2022}
\abst{\myAbstract}
\kword{\myKeywords}
\begin{document}
\maketitle

\else

\documentclass[3p,letter]{elsarticle}
\usepackage{graphicx}
\usepackage{amsmath}
\usepackage{amssymb}
\usepackage{textcomp}
\usepackage{xcolor}
\definecolor{myblue}{rgb}{0.0, 0.0, 0.6}
\usepackage{hyperref}
\hypersetup{
  colorlinks = true,
  citecolor  = myblue,
  linkcolor  = myblue,
  urlcolor   = myblue
}
\journal{
  The Proceedings of the 24th International Spin Symposium (SPIN2021)}
\begin{document}
\begin{frontmatter}
\title{\myTitle}
\author[1]{A.~A. Poblaguev}\ead{poblaguev@bnl.gov}
\author[1]{G. Atoian}
\author[2]{N.~H. Buttimore}
\author[1]{A. Zelenski}
\address[1]{Brookhaven National Laboratory, Upton, New York 11973, USA}
\address[2]{School of Mathematics, Trinity College Dublin, Dublin 2, Ireland}
\date{December 12, 2022}%

\begin{abstract} \myAbstract \end{abstract}
\end{frontmatter}

\fi

\section{Introduction}

The physics program at the future Electron Ion Collider (EIC) \cite{Accardi:2012qut} requires precision measurement of the proton and helion (${}^3\rm{He}$) beam absolute polarization with systematic uncertainties of about\,\cite{AbdulKhalek:2021gbh}
\begin{equation}
  \sigma_P^\text{syst}/P \lesssim 1\%.
  \label{eq:systEIC}
\end{equation}

For high energy, $\sim$100\,GeV, polarized proton beams, such an accuracy was already achieved at the Relativistic Heavy Ion Collider (RHIC). At RHIC, the absolute vertical proton beam polarization is determined by the Atomic Polarized Hydrogen Gas Jet Target (HJET)\,\cite{Zelenski:2005mz}. The vertically directed jet target polarization, $P_\text{jet}\!=\!96\!\pm\!0.1\%$, is monitored by a Breit-Rabi polarimeter and is flipped every 5\,minutes.  A critically important feature for the HJET polarimeter is a relatively low-density gas jet target with no walls or windows which allows one to make continuous measurements in the Coulomb-nuclear interference (CNI) region.

\begin{figure}[t]
  \begin{minipage}[t]{0.48\columnwidth}
  \begin{center}
       \option
       \includegraphics[width=0.9\textwidth]{Fig01_HjetView.eps}
       \else    
       \includegraphics[width=0.9\textwidth]{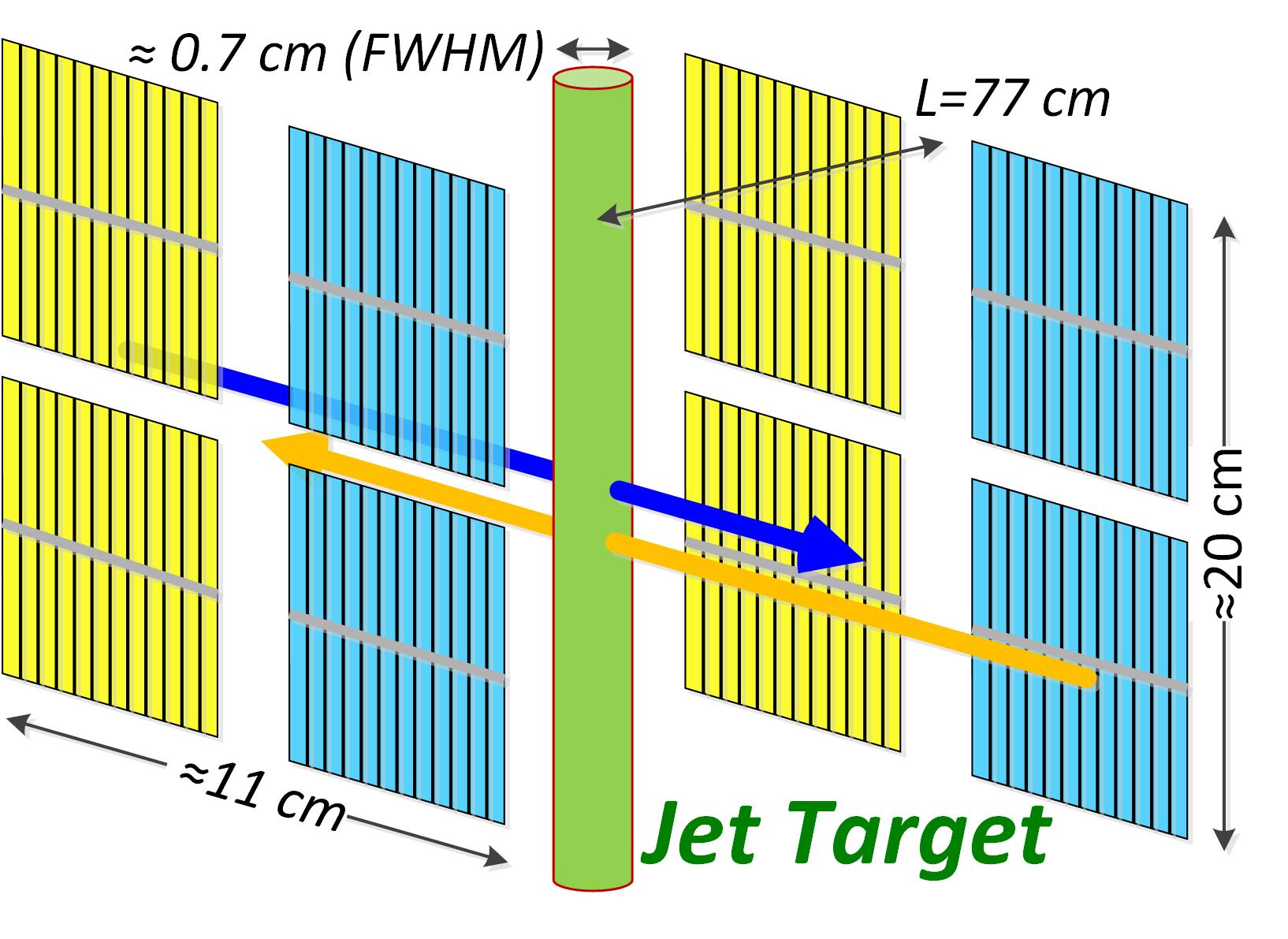}
       \fi
  \end{center}
  \caption{\label{fig:HJET}
   Schematic view of the HJET recoil spectrometer. Eight detectors with vertically oriented Si strips are used to measure the recoil proton time, amplitude, and recoil angle, concurrently for both RHIC beams.
  }
  \end{minipage} \hfill \begin{minipage}[t]{0.48\columnwidth}
   \begin{center}
       \option
       \includegraphics[width=0.9\textwidth]{Fig02_StripTR.eps}
       \else
       \includegraphics[width=0.9\textwidth]{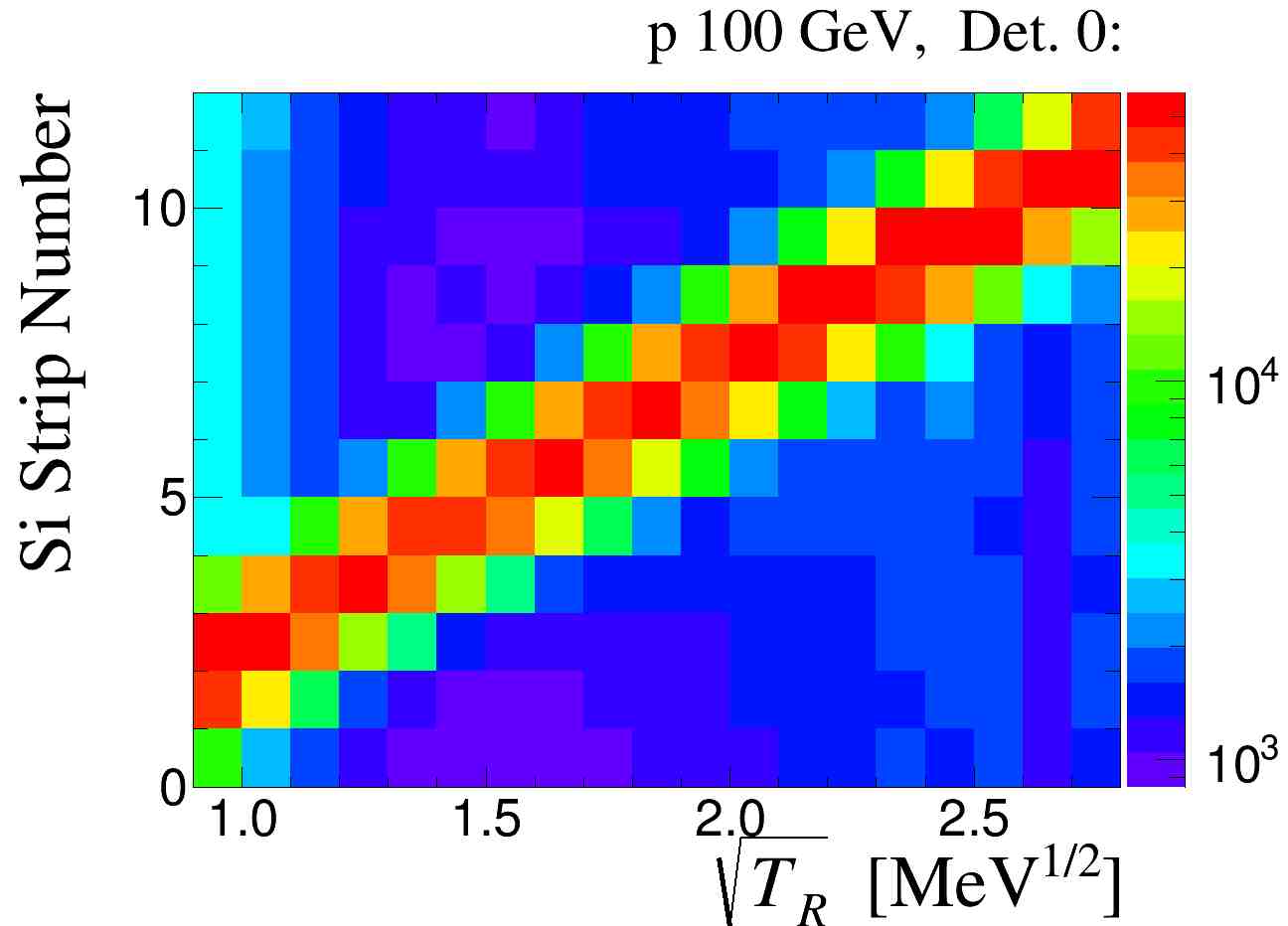}
       \fi
     \end{center}
      \caption{\label{fig:Strips}
        Correlation between Si strip number (in a detector) and the recoil kinetic energy $T_R$ allows one to isolate elastic events. Since the background rate is almost independent of the Si strip number (for given $T_R$)\,\cite{Poblaguev:2020qbw}, the background events can be subtracted from the elastic data.
    }
    \end{minipage}
\end{figure}

During RHIC run, the stored beams with alternatively polarized bunches are scattered on the hydrogen jet target and recoil protons are counted in the left/right symmetric recoil spectrometer detectors at 90 degrees to the beam direction\,\cite{Poblaguev:2020qbw}.  The spectrometer is depicted in Fig.\,\ref{fig:HJET}. Isolation of the elastic events is explained in Fig.\,\ref{fig:Strips}.

For elastic $\mathit{pp}$ scattering, the concurrently measured beam and jet spin correlated asymmetries
\begin{equation}
        a_\text{beam}=\langle A_\text{N}\rangle P_\text{beam},\quad%
        a_\text{jet} =\langle A_\text{N}\rangle P_\text{jet},\qquad\to\qquad%
        P_\text{beam} = P_\text{jet}\times\left(a_\text{beam}/a_\text{jet}\right)
        \label{eq:BeamPol}
\end{equation}
allow one to determine the beam polarization $P_\text{beam}$ without knowing the analyzing power $A_\text{N}$. At HJET, the asymmetries can be measured as functions of the recoil proton energy $0.6\!<\!T_R\!<\!10.6\,\text{MeV}$ which was used for experimental evaluation of analyzing power dependence $A_\text{N}(t)$ on momentum transfer $t\!=\!-2m_pT_R$. A detailed description of the HJET data analysis, including subtraction of a background and evaluation of systematic uncertainties, is given in Ref.\,\cite{Poblaguev:2020qbw}.  

In previous RHIC polarized proton Runs 15 ($E_\text{beam}\!=\!100\,\text{GeV}$) and 17 (255\,GeV), the beam polarization of about $P_\text{beam}\!\sim\!55\%$ was measured with systematic error of $\sigma^\text{syst}_P/P\!\lesssim\!0.5\%$ and typical statistical uncertainties of $\sigma_P^\text{stat}\!\sim\!2\%$ per 8\,hour RHIC store\,\cite{Poblaguev:2020qbw}. Also, the elastic ${pp}$ analyzing power was precisely determined at both beam energies\,\cite{Poblaguev:2019saw}. 

Since 2015, HJET has routinely operated in the RHIC ion beams. The recoil spectrometer performance was found to be very similar to that of a proton beam and is very  stable for a wide range of ion species ($d$, Al, Zr, Ru, Au) and over a wide range, 3.8\,--\,100 GeV/n, of the beam energies used\,\cite{Poblaguev:2020qbw}.

Here, we investigate a possibility to adapt the HJET method for a precise measurement of the ${}^3\text{He}$ beam polarization at EIC.

\section{${}^3\rm{He}$ beam polarimetry at EIC}

To straightforwardly implement the HJET method (\ref{eq:BeamPol}) for absolute helion polarimetry at EIC, one should employ a polarized ${}^3\rm{He}^\uparrow$ target. But, for such a target it may be an issue to satisfy the EIC requirements, in particular, high polarization, which can be monitored with $\ll\!1\%$ accuracy, and consistency with recoil helion measurements in the CNI region. For example, ${}^3\rm{He}$ gas can be polarized up to $\sim$70\% in a cell and, then, it can be injected into the collision chamber producing a jet-like target. However, confirmed polarization uncertainty for such a ${}^3\rm{He}$ jet is
$\sigma_P/P\!=\!3.4\%$\,\cite{DeSchepper:1998gc}. Some other obviously anticipated concerns about the method adaption are: ({\em i}) the recoil helion kinematics essentially differs from that of the proton case; ({\em ii}) possible breakup of the beam and/or target may result in uncontrolled backgrounds; ({\em iii}) the EIC bunch spacing (10 ns) will be much shorter compared to that (107 ns) in HJET measurements at RHIC. A possibility to overcome the mentioned potential problems is being investigated by RHIC Polarimetry Group.

Here, we will consider HJET feasibility to determine the EIC ${}^{3}\rm{He}$ beam polarization with required accuracy (\ref{eq:systEIC}). For such a measurement, the $h^\uparrow{p}$ and $p^\uparrow{h}$ analyzing powers do not cancel in Eq.\,(\ref{eq:BeamPol}). Thus, the ratio $A_\text{N}^{hp}(t)/A_\text{N}^{ph}(t)$ must be predetermined. Also we will consider the effects of the helion beam breakup and shorter bunch spacing at EIC.

\section{The analyzing power}

For high energy forward elastic $\mathit{p^\uparrow{p}}$ scattering, the analyzing power structure is theoretically well understood\,\cite{Kopeliovich:1974ee,Buttimore:1978ry,Buttimore:1998rj}. Neglecting some small, well determined corrections,
\begin{equation}
A_N(t) = \frac{\sqrt{-t}}{m_p}\,%
       \frac{\kappa_p-2I_5 -2R_5\,t/t_c}{t_c/t-2(\rho+\delta_C)+t/t_c},
\end{equation}
where $\kappa_p\!=\!\mu_p\!-\!1\!=\!1.793$ is anomalous magnetic moment of a proton, $\rho$ is forward real to imaginary ratio, $\delta_C\!\sim\!0.02$ is Coulomb phase\,\cite{Cahn:1982nr,Kopeliovich:2000ez}, and $-t_c=8\pi\alpha/\sigma_\text{tot}\!\approx\!0.002\,\text{GeV}^2$ is expressed via total $\mathit{pp}$ cross section. The dominant term, $\kappa_p$ in the numerator is defined by interference of the electromagnetic spin flip and hadronic non-flip amplitudes. Corrections due to the hadronic spin flip amplitude, parameterized by $r_5\!=R_5\!+\!I_5$\,\cite{Buttimore:1998rj}, are of about 2\%\,\cite{Poblaguev:2019saw}. For, elastic $\mathit{pp}$ scattering, the denominator, which is proportional to the differential cross section,  is precisely known from numerous experimental studies, but this is not the case for proton-helion scattering.

However, the $\mathit{hp}$ cross section uncertainties cancel in the ratio of the  $h^\uparrow{p}$ and $p^\uparrow{h}$ analyzing powers:
\begin{equation}
  {\cal R}=\frac{A_\text{N}^{hp}(t)}{A_\text{N}^{ph}(t)}
   = \frac{\kappa_h-2I_5^{hp} -2R_5^{hp}\,t/t_c}{\kappa_p-2I_5^{ph} -2R_5^{ph}\,t/t_c},
\end{equation}
where $k_h\!=\!\mu_h/2\!-\!1/3\!=\!-1.398$ is derived from the magnetic moment of a helion\,\cite{Buttimore:2009zz}. Although small, the hadronic spin-flip amplitude should be considered when achieving the required (\ref{eq:systEIC}) accuracy of the beam polarization measurements at EIC.

Using proton-proton $r_5^{pp}$, precisely determined at HJET\,\cite{Poblaguev:2019saw}, one can evaluate $r_5^{ph}$ and $r_5^{hp}$ with sufficient accuracy. In Ref.\,\cite{Kopeliovich:2000kz}, it was shown that the $p^\uparrow\rm{A}$ value of $r_5$ can be well approximated by the $p{}^\uparrow{}p$ one, $r_5^{ph}\!=\!r_5^{pp}$. Since ${}^3\rm{He}$ is carried mostly by the neutron, $r_5^{hp}\!\approx\!r_5^{pp}/3$\,\cite{Buttimore:2001df}. Assuming $|r_5^{pp}|\lesssim0.02$, model dependent uncertainties in such extrapolation of $r_5^{pp}$ to $r_5^{ph}$ and $r_5^{hp}$ can be evaluated\,\cite{Poblaguev:2022gqy} as
\begin{equation}
        r_5^{ph}=r_5^{pp}\pm0.001\pm 0.001\:\!i,\qquad
        r_5^{hp}=0.27r_5^{pp}\pm0.001\pm 0.001\:\!i.
\end{equation}

\section{The helion beam breakup}

Since only recoil protons are detected at HJET, possible breakup of a beam helion, $h\to d+p$ can contaminate the elastic ${hp}$ scattering data. Generally, the total breakup cross section can be much larger than the elastic one, e.g. in deuteron-proton scattering\,\cite{Glagolev:2006nv}. However, the differential breakup cross section must vanish if $t\!\to\!0$, unless the beam particle can spontaneously decay. Therefore, for forward scattering events, $|t|\!<\!0.02\,\text{GeV}^2$ ($T_R\!<\!10\,\text{MeV}$), which can be detected at HJET, the breakup (or quasi-elastic, {\em qel}) fraction, $f\!=\!\sigma_\text{qel}\,/\,\sigma_\text{el}\big|_\text{HJET}$, is expected to be small. In a recent evaluation of this value for a beam of Au, it was found that $f_\text{Au}\!\lesssim1\%$.

To evaluate the analyzing power ratio $\cal{R}$ dependence on the breakup, one should substitute the (unpolarized) hadronic  elastic amplitude by a sum of elastic and quasi-elastic amplitudes $\phi_+(t)\to\phi_+(t)+\sum{\phi_\text{qel}(t,\Delta)}=\phi_+(t)\times[1+\tilde{\omega}(t)]$, where $\Delta=(M_X^2-m_h^2)/2m_h$ is the breakup mass excess. Using Run 16 deuteron beam (10--100\,GeV/n) measurements at HJET, it was calculated that $\left|\tilde{\omega}(t)\right|\!<\!0.05$ (upper limit)\,\cite{Poblaguev:2022hsi} for the helion beam.  Thus,
\begin{equation}
  {\cal R}_\text{el} \to {\cal R}_\text{el+qel}=\frac{\kappa_h\times[1+\tilde{\omega}(t)]-2I_5^{hp} -2R_5^{hp}\,t/t_c}
        {\kappa_p\times[1+\tilde{\omega}(t)]-2I_5^{ph} -2R_5^{ph}\,t/t_c} \approx {\cal R}_\text{el}.
\label{eq:R_corr}
\end{equation}
Here we assume that electromagnetic amplitudes are about the same for elastic and quasi-elastic scattering (if $t\to0$). Also, we neglected the breakup corrections to hadronic spin flip amplitudes since $|r_5|\!\sim\!0.02$ is small. In the approximation used, the breakup corrections cancel in the ratio\,(\ref{eq:R_corr}).

\section{The EIC bunch spacing}

\begin{figure}[t]
  \begin{minipage}[t]{0.62\columnwidth}
  \begin{center}
    \option
    \includegraphics[width=0.48\textwidth]{Fig03a_rhicAT_0.eps}\hfill%
    \includegraphics[width=0.48\textwidth]{Fig03b_rhicAT.eps} 
    \else
    \includegraphics[width=0.48\textwidth]{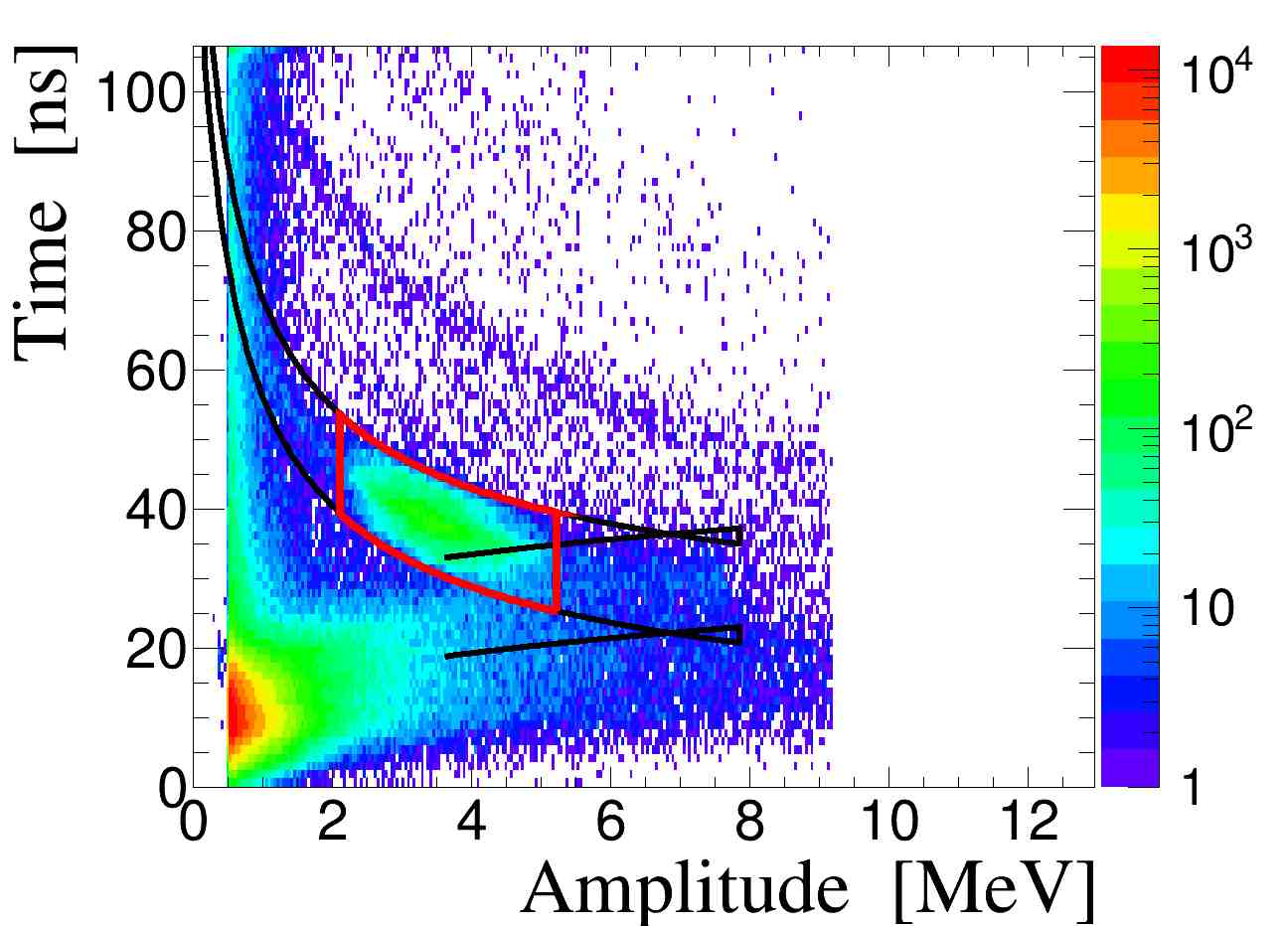}\hfill%
    \includegraphics[width=0.48\textwidth]{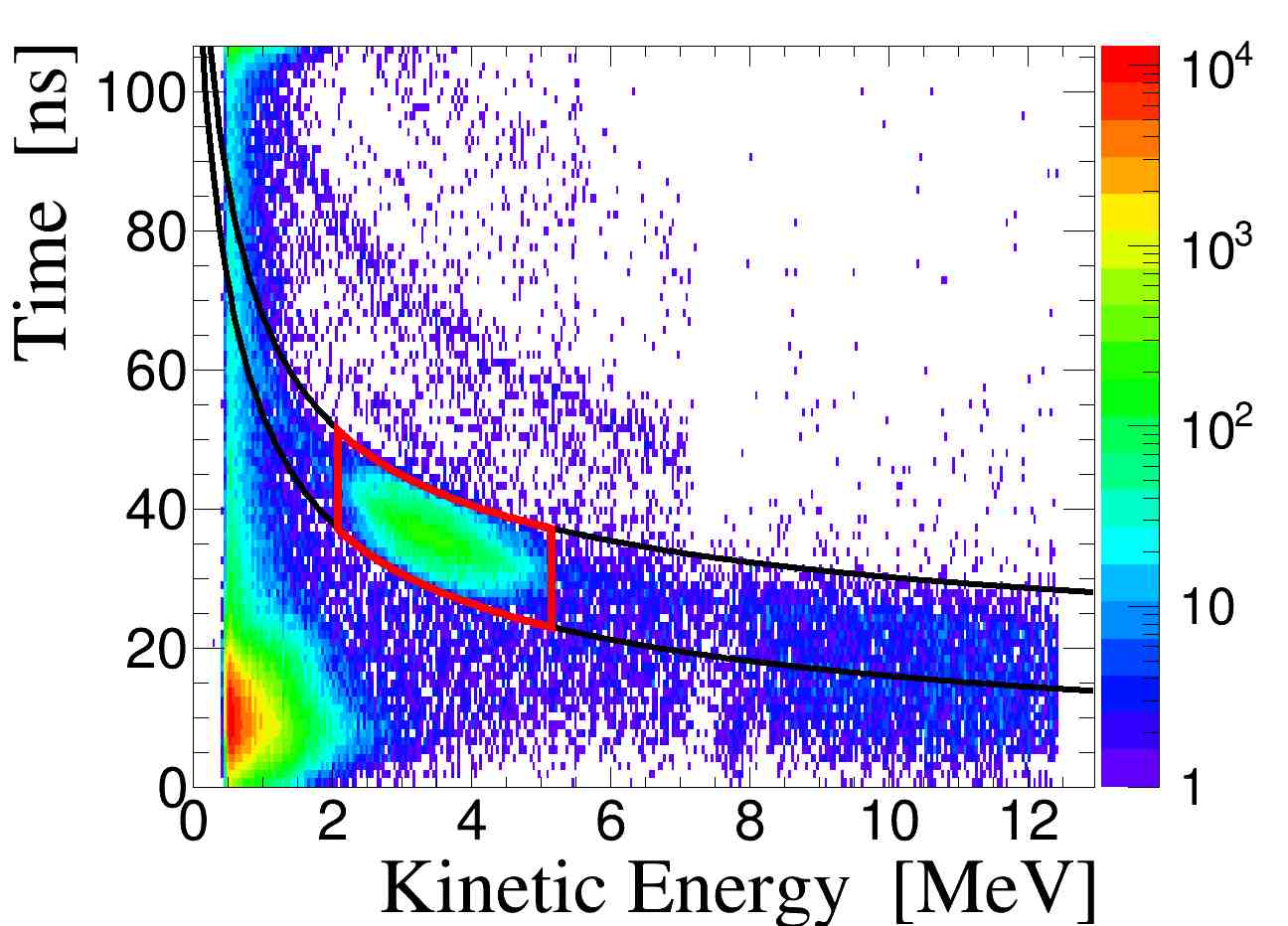} 
    \fi
  \end{center}
  \caption{\label{fig:rhicAT}
    Left: a typical detected signal time/amplitude distribution in a Si strip. Black lines specify the time/amplitude area corresponding to the recoil protons. Red contours isolate the elastic $\mathit{pp}$ events. Right: the same distribution after reconstruction of kinetic energy of the punch-trough protons. 
  }
  \end{minipage} \hfill \begin{minipage}[t]{0.36\columnwidth}
   \begin{center}
      \option
      \includegraphics[width=0.844\textwidth]{Fig04_pow150V.eps}
      \else
      \includegraphics[width=0.844\textwidth]{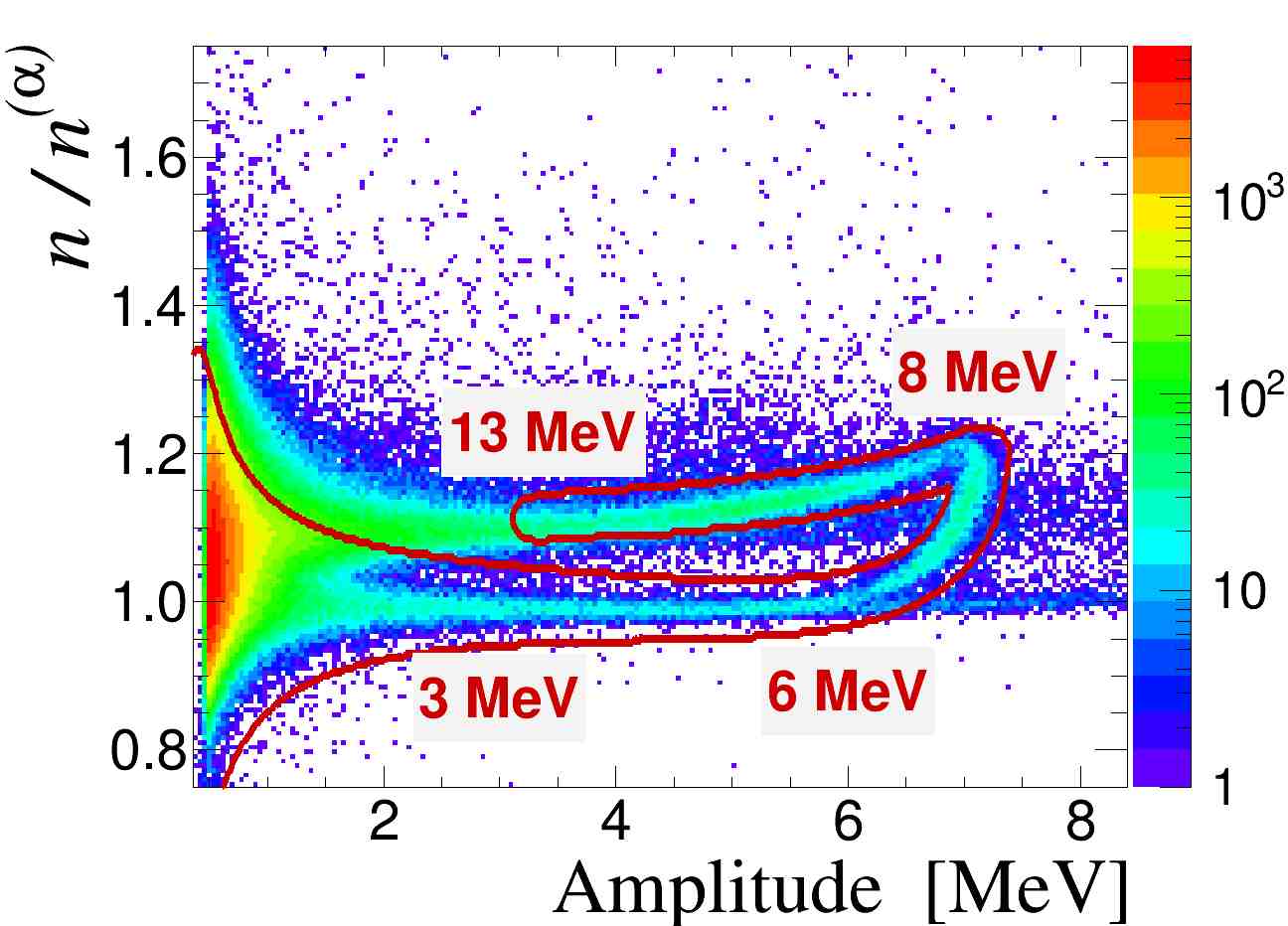}
      \fi
    \end{center}
      \caption{\label{fig:pow}
    Separation of the stopped and punch-trough recoil protons based on the correlation of the signal waveform parameters $n$ and $A$.
    }
    \end{minipage}
\end{figure}

For HJET measurements at RHIC, a typical time/amplitude distribution in a Si strip is shown in Fig.\,\ref{fig:rhicAT}\,(left). The data rate is strongly dominated by {\em prompts}, characterized by low deposited energy and small time of flight (TOF).  The recoil protons with energy above $7.8\,\text{MeV}$ punch through a Si strip and, thus, only part of the kinetic energy is detected. Such events as well as {\em prompts} might be a serious problem for beam polarization  measurement at EIC. Due to the reduced, 10\,ns, bunch spacing at EIC, the elastic $\mathrm{pp}$ signals might be mismatched with punch-trough protons and background events (in particular, {\em prompts}) from other bunches. This is anticipated to be a common problem for proton and helion beam polarimetry at EIC.

Assuming that {\em prompts} are fast particles penetrating the Si strip, two possible solutions of the problem were considered.

Since the recoil proton signal wave form shape parametrization, $w(t)\propto At^n\,\exp{(-t/\tau)}$, used in the data analysis, depends on kinetic energy $T_R$ of the proton\,\cite{Poblaguev:2020qbw} (see Fig.\,\ref{fig:pow}), $T_R$ for punch-through protons can be reconstructed from the measured amplitude as shown in Fig.\,\ref{fig:rhicAT}\,(right). Such a transformation significantly reduces mismatching signals from different bunches.

Using RHIC Run 17 (255 GeV protons) data, HJET performance for 8.9 ns bunch spacing was emulated\,\cite{Poblaguev:2020Og}. For that, the measured time of each event was shifted $t \to t + k\tau/12$, where $k$ is randomly chosen from $k\in(-1.5, -0,5, 0.5, 1.5)$ and $\tau\!=\!106.6\,\text{ns}$ is the bunch spacing in Run\,17. This transformation approximates the proposed bunch splitting into four at EIC. In addition, every event was triplicated by the time shifts $\pm\tau/3$.

Such emulated data was processed using regular HJET data analysis software.
It was found that, for $T_R\!>\!2\,\text{MeV}$, the EIC bunch spacing will not alter, within $|\delta{}P/P|\!\lesssim\!0.3\%$ uncertainty, the measured beam polarization.

   Anther approach to solve the problem is to use double-layer Si detectors to veto {\em prompts}\,\cite{Poblaguev:2020Og}. One of the HJET detectors was replaced with a double layer prototype.  The measurement was done using a beam of Au, which was available at that time. The efficiency of {\em prompt} vetoing appeared to be only about $\sim$50\%. For {\em prompts} which did not trigger the second layer, it was found that time of flight is about the same as for a photon and the signal wave form shape is similar to that of stopped recoil protons. The result obtained disagrees with an initial assumption about the {\em prompts}. Currently we have no satisfactory numerical model for the {\em prompt} signals.

Nonetheless, we are positive about HJET feasibility to operate in the EIC 10 ns bunch spacing beam (if the recoil proton energy will be limited to $T_R\!>\!2\,\text{MeV}$ in the data analysis).

\section{Summary}
Based on the analysis above, we should conclude that the absolute ${}^3\rm{He}$ beam polarization at EIC can be determined with the required precision (\ref{eq:systEIC}). For that, we suggest measuring the polarization as a function of the recoil proton energy: 
\begin{equation}
  P_m(T_R) = P_\text{jet} \frac{a_\text{beam}^{(h)}(T_R)}{a_\text{jet}^{(p)}(T_R)}\times%
  \frac{\kappa_p-m_p^2/m_hE_\text{beam}-2I_5-2R_5\,T_R/T_c}%
       {\kappa_h-m_h/E_\text{beam}    -0.54I_5-0.54R_5\,T_R/T_c}.
       \label{eq:He3_Pol}
\end{equation}
Here, $\kappa_p=1.793$, $\kappa_h=-1.398$, $T_c=4\pi\alpha/m_p\sigma_\text{tot}^{hp}\approx0.7\,\text{MeV}$,
and the proton-proton $r_5=R_5+iI_5$ is
predetermined for the same beam energy $E_\text{beam}$ (in GeV/n units). Eq.\,(\ref{eq:He3_Pol})
also includes an energy dependent correction to the 
electromagnetic amplitude\,\cite{Poblaguev:2019vho}, which was not discussed above. Extrapolating the $P_m(T_R)$ dependence to $T_R\to0$, the helion beam polarization can be found: 
\begin{equation}
 \!\!P_\text{beam}=P_m(0)\times%
  \left[1\!\pm\!0.006_\text{syst}\!\pm\!0.005_{r_5}\!\pm\!0.002_\text{mod}\right].
  \label{eq:PhErrs}
\end{equation}
The specified uncertainties are estimates of systematic errors in HJET measurements (syst), in the value of $r_5$, and in the model dependent derivation of Eq.\,(\ref{eq:He3_Pol}).

It should be pointed out that HJET (with possible improvements if required) is being considered for measurement of the proton beam absolute polarization at EIC. Here, we advocate that exactly that polarimeter also has the ability to precisely measure helion beam polarization (if proton-proton $r_5$ for the beam energy used is already known).

However, development of alternative methods of measuring the helion beam polarization, (e.g. using polarized ${}^3\rm{He}$ target) is very important to obtain a confident result.
Also, it may be crucially helpful for the data analysis to extend the measurement to lower kinetic energies $T_R<2\,\text{MeV}$.

\option

\else 

\providecommand{\href}[2]{#2}\begingroup\raggedright\endgroup

\fi

\end{document}